\documentclass[preprintnumbers,amsmath,amssymbm,prd]{revtex4}
\usepackage{epsfig}
\usepackage{graphicx}
\usepackage{amssymb}

\begin{document}

\title{Upper bound on the gravitational masses of stable spatially regular charged compact objects}
\author{Shahar Hod}
\affiliation{The Ruppin Academic Center, Emeq Hefer 40250, Israel}
\affiliation{ } \affiliation{The Hadassah Institute, Jerusalem
91010, Israel}
\date{\today}

\begin{abstract}

\ \ \ In a very interesting paper, Andr\'easson has recently proved
that the gravitational mass of a spherically symmetric compact
object of radius $R$ and electric charge $Q$ is bounded from above
by the relation
$\sqrt{M}\leq{{\sqrt{R}}\over{3}}+\sqrt{{{R}\over{9}}+{{Q^2}\over{3R}}}$.
In the present paper we prove that, in the dimensionless regime
${{Q}/{M}}<\sqrt{{9/8}}$, a stronger upper bound can be derived on
the masses of physically realistic ({\it stable}) self-gravitating
horizonless compact objects: $M<{{R}\over{3}}+{{2Q^2}\over{3R}}$.
\end{abstract}
\bigskip
\maketitle

\section{Introduction}

The asymptotically measured gravitational mass $M$ of a spherically
symmetric asymptotically flat Schwarzschild black-hole spacetime is
directly related to the horizon radius $R$ by the simple relation
$M=R/2$ \cite{Chan,Noteunits}. It is well known that a stronger
upper bound on the gravitational masses of spatially regular
self-gravitating horizonless compact objects is provided by the
physically important Buchdahl bound $M\leq 4R/9$ \cite{Buch}.

Similar bounds are known to exist for charged self-gravitating
compact objects. In particular, charged Reissner-Nordstr\"om black
holes are characterized by the simple relation $M=R/2+Q^2/2R$
\cite{Chan}. In a physically interesting paper, Andr\'easson
\cite{Hak1} has recently derived the stronger upper bound
\begin{equation}\label{Eq1}
\sqrt{M}\leq{{\sqrt{R}}\over{3}}+\sqrt{{{R}\over{9}}+{{Q^2}\over{3R}}}\
\end{equation}
on the gravitational masses of spatially regular horizonless charged compact objects.

In the present paper we raise the following physically intriguing
question: Can one improve the important upper bound (\ref{Eq1}) on
the masses of self-gravitating charged compact objects by adding to
the characteristic properties of the corresponding horizonless
curved spacetimes the physically motivated requirement of dynamical
{\it stability}?

As we shall explicitly show below, the above stated question is
directly related to the physically important theorem presented
recently in \cite{CBH} (see also \cite{Hodrw,Hoddd}), according to
which the innermost null circular geodesic of an horizonless compact
object, if it exists, is stable \cite{Notesth}. In particular,
combining this interesting physical property of the spatially
regular self-gravitating compact objects that we consider in the
present paper with the intriguing assertion made in \cite{Keir} (see
also \cite{CarC,Hodt1}), according to which horizonless spacetimes
which possess stable null circular geodesics (stable closed light
rings) are expected to develop non-linear instabilities in response
to the presence of time-dependent massless perturbation fields
\cite{Notekk}, one concludes that spatially regular compact objects
that possess light rings in their exterior spacetime regions are
dynamically unstable.

Motivated by the physically important observations made in
\cite{CBH,Keir} regarding the (in)stability properties of
horizonless compact objects, in the present paper we shall use {\it
analytical} techniques in order to derive an improved upper bound
[see Eq. (\ref{Eq22}) below] on the maximally allowed gravitational
masses $M^{\text{max}}(R,Q)$ of dynamically {\it stable} spatially
regular charged compact objects.

\section{Description of the system}

We consider self-gravitating horizonless charged compact objects
whose spatially regular curved spacetimes are described by the
spherically symmetric line element
\cite{Chan,ShTe,CarC,Hodt1,Hodt2,Notesc}
\begin{equation}\label{Eq2}
ds^2=-e^{-2\delta}\mu dt^2 +\mu^{-1}dr^2+r^2(d\theta^2 +\sin^2\theta
d\phi^2)\  .
\end{equation}
The radially dependent metric functions $\mu=\mu(r)$ and
$\delta=\delta(r)$ are related to the composed energy-momentum
tensor $T^{\mu}_{\nu}(\text{total})=T^{\mu}_{\nu}(\text{matter})+
T^{\mu}_{\nu}(\text{electromagnetic-field})$ of the charged matter
configurations by the Einstein field equations \cite{Chan}
\begin{equation}\label{Eq3}
G^{\mu}_{\nu}=8\pi[T^{\mu}_{\nu}(\text{matter})+T^{\mu}_{\nu}(\text{electromagnetic-field})]\
,
\end{equation}
which, using the curved line element (\ref{Eq2}) and the functional
expressions \cite{BekMay}
\begin{equation}\label{Eq4}
{T^{\text{em}}}^{t}_{t}={T^{\text{em}}}^{r}_{r}=
-{T^{\text{em}}}^{\theta}_{\theta}=-{T^{\text{em}}}^{\phi}_{\phi}=-{{Q^2(r)}\over{8\pi
r^4}}\
\end{equation}
for the components of the electromagnetic (em) energy-momentum
tensor, can be expressed in the differential forms
\cite{BekMay,Hodt1,Noteprm}
\begin{equation}\label{Eq5}
\mu'=-8\pi r\Big[\rho+{{Q^2(r)}\over{8\pi
r^4}}\Big]+{{1-\mu}\over{r}}\
\end{equation}
and
\begin{equation}\label{Eq6}
\delta'=-{{4\pi r(\rho +p)}\over{\mu}}\  .
\end{equation}
Here $Q(r)$ is the electric charge contained within a sphere of
areal radius $r$ \cite{BekMay}, $\rho\equiv
-T^{t}_{t}(\text{matter})$, and $p\equiv T^{r}_{r}(\text{matter})$
\cite{Bond1}.
%Following \cite{Hak1}, we shall assume that the radial
%pressure of the matter fields is non-negative:
%\begin{equation}\label{Eq7}
%p\geq0\  .
%\end{equation}

The metric functions $\{\mu,\delta\}$ of the horizonless spatially
regular asymptotically flat spacetime (\ref{Eq2}) are respectively
characterized by the near-origin and far-region functional relations
\cite{Hodt1}
\begin{equation}\label{Eq7}
\mu(r\to 0)\to1\ \ \ \ ; \ \ \ \ \mu(r\to\infty)\to1\
\end{equation}
and \cite{Hodt1}
\begin{equation}\label{Eq8}
\delta(0)<\infty\ \ \ \ ; \ \ \ \ \delta(r\to\infty)\to 0\ .
\end{equation}
In particular, the Einstein equation (\ref{Eq5}) implies that the
radial metric function $\mu(r)$ can be expressed in the
mathematically compact form \cite{Hak1}
\begin{equation}\label{Eq9}
\mu(r)=1-{{2m(r)}\over{r}}+{{Q^2(r)}\over{r^2}}\  ,
\end{equation}
where $m(r)$ is the gravitational mass contained within a sphere of
radius $r$ \cite{BekMay,Hak1}. For later purposes we note that, as
explicitly proved in \cite{Hodt1}, regular self-gravitating
matter configurations with asymptotically measured finite ADM masses
are characterized by the asymptotically decaying functional behavior
\begin{equation}\label{Eq10}
r^3 T^{r}_{r}(\text{total})\to 0\ \ \ \ \text{for}\ \ \ \
r\to\infty\ .
\end{equation}

\section{The upper bound on the gravitational masses of stable spatially regular horizonless charged compact
objects}

In the present section we shall prove that, in the dimensionless
regime  \cite{Noteqq,Notets}
\begin{equation}\label{Eq11}
{{Q}\over{M}}\leq \sqrt{{{9}\over{8}}}\  ,
\end{equation}
one can use the instability properties of spatially regular
horizonless spacetimes which possess light rings
\cite{CBH,Hodrw,Hoddd,Keir}, in order to derive an upper bound on
the gravitational masses of physically realistic ({\it stable})
charged compact objects. In particular, below we shall explicitly
show that the newly derived bound [see Eq. (\ref{Eq22}) below] is
stronger than the important upper bound (\ref{Eq1}).

The functional equation which determines the radii of light rings in
the curved spacetime (\ref{Eq2}) was derived in
\cite{Chan,CarC,Hodt1}. For completeness of the presentation, we
shall first provide a brief sketch of the analytical derivation of
the functional relation which characterizes the null circular
geodesics of the charged spacetime. As explicitly shown in
\cite{Chan,CarC}, the circular null trajectories which characterize
the spherically symmetric spacetime (\ref{Eq2}) are determined by
the two relations \cite{Notethr,Notedot}
\begin{equation}\label{Eq12}
V_r=E^2\ \ \ \ \text{and}\ \ \ \ V'_r=0\  ,
\end{equation}
where the effective potential $V_r$ is given by the functional
expression \cite{Chan,CarC,Hodt1}
\begin{equation}\label{Eq13}
E^2-V_r\equiv \dot
r^2=\mu\Big({{E^2}\over{e^{-2\delta}\mu}}-{{L^2}\over{r^2}}\Big)\  .
\end{equation}
Here the energy $E$ and the angular momentum $L$ are conserved
quantities which reflect the fact that the metric components of the
spherically symmetric static spacetime (\ref{Eq2}) are independent of
the time and angular coordinates $\{t,\phi\}$
\cite{Chan,CarC,Hodt1}.

Substituting Eq. (\ref{Eq13}) into Eq. (\ref{Eq12}) and using the
Einstein differential equations (\ref{Eq5}) and (\ref{Eq6}), one
finds that the light rings of the spherically symmetric static
curved spacetime (\ref{Eq2}) are determined by the compact
functional relation
\begin{equation}\label{Eq14}
{\cal R}(r)\equiv 3\mu-1-8\pi r^2\Big[p-{{Q^2(r)}\over{8\pi
r^4}}\Big]=0\ \ \ \ \text{for}\ \ \ \ r=r_{\gamma}\  .
\end{equation}
In addition, taking cognizance of Eqs. (\ref{Eq7}), (\ref{Eq10}),
and (\ref{Eq14}), one deduces that the dimensionless function ${\cal
R}(r)$, whose zeroes determine the discrete radii of the null
circular geodesics of the spherically symmetric spacetime
(\ref{Eq2}), is characterized by the two boundary relations
\begin{equation}\label{Eq15}
{\cal R}(r=0)=2\ \ \ \ \text{and}\ \ \ \ {\cal R}(r\to\infty)\to 2\
.
\end{equation}
These simple relations imply that spatially regular horizonless
compact objects are generally characterized by an {\it even} number of
null circular geodesics \cite{CBH,Hoddd,NoteHoddd}.

The stability properties of the null circular geodesics are
generally determined by the second spatial derivative of the
effective curvature potential (\ref{Eq13}) \cite{Chan,CarC}. In
particular, unstable light rings are characterized by locally
concave radial potentials with $V''_r(r=r_{\gamma})<0$, whereas
stable circular geodesics which, as discussed in \cite{Keir}, are
associated with non-linear instabilities of the corresponding curved
spacetimes, are characterized by locally convex curvature potentials
with the property $V''_r(r=r_{\gamma})>0$ \cite{Chan,CarC}. Taking
cognizance of Eqs. (\ref{Eq5}), (\ref{Eq6}), (\ref{Eq12}), and
(\ref{Eq13}), and using the conservation equation $T^{\mu}_{r
;\mu}=0$ \cite{BekMay}, one finds the simple functional relation
\cite{Hoddd,Hodrw}
\begin{equation}\label{Eq16}
V''_r(r=r_{\gamma})=-{{E^2e^{2\delta}}\over{\mu r_{\gamma}}}\times
{\cal R}'(r=r_{\gamma})\  .
\end{equation}
%where [see Eqs. (\ref{Eq5}) and (\ref{Eq15})]
%\begin{equation}\label{Eq18}
%{\cal R}'(r=r_{\gamma})={{2}\over {r_{\gamma}}}\big[1-8\pi
%r^2_{\gamma}(\rho+p_T)\big]\  .
%\end{equation}

From Eqs. (\ref{Eq14}) and (\ref{Eq15}) one deduces that the
innermost null circular geodesic, $r=r^{\text{innermost}}_{\gamma}$,
of a spatially regular horizonless compact object is generally
\cite{NoteHoddd} characterized by the properties
\begin{equation}\label{Eq17}
{\cal R}(r=r^{\text{innermost}}_{\gamma})=0\ \ \ \ \text{and}\ \ \ \
{\cal R}'(r=r^{\text{innermost}}_{\gamma})<0\  .
\end{equation}
In particular, the innermost light ring of a spatially regular
compact object, if it exists, is generally {\it stable} with the
property $V''_r(r=r^{\text{innermost}}_{\gamma})>0$ [see Eqs.
(\ref{Eq16}) and (\ref{Eq17})] \cite{Hodrw,CBH,Hoddd}.

The exterior spacetime regions $(r\geq R)$ of the spherically
symmetric charged compact objects that we consider in the present
paper are characterized by the relations \cite{Hak1}
\begin{equation}\label{Eq18}
\rho=p=0\
\end{equation}
and \cite{Chan}
\begin{equation}\label{Eq19}
\mu(r)=1-{{2M}\over{r}}+{{Q^2}\over{r^2}}\ \ \ \ \text{for}\ \ \ \
r\geq R\  ,
\end{equation}
where $\{M,Q\}$ are respectively the total gravitational mass and
the total electric charge of the spherically symmetric spacetime as
measured by asymptotic observers.

Let us assume that the spatially regular charged compact object
possesses an external light ring with $r_{\gamma}>R$. Substituting
Eqs. (\ref{Eq18}) and (\ref{Eq19}) into the functional relation
(\ref{Eq14}), which characterizes the null circular geodesics of the
spherically symmetric spacetime (\ref{Eq2}), one finds the
remarkably simple expression
\begin{equation}\label{Eq20}
r^{\text{outer}}_{\gamma}={1\over2}\big(3M+\sqrt{9M^2-8Q^2}\big)
\ \ \ \ \ \text{for}\ \ \ \ \ {{Q}\over{M}}\leq \sqrt{{9\over8}}\
\end{equation}
for the radius of the outer light ring.

As discussed above, the presence of the light ring (\ref{Eq20})
outside the surface of a spatially regular horizonless compact
object \cite{Notepr} implies the existence of a second (stable)
light ring (with the property
$r^{\text{innermost}}_{\gamma}<r^{\text{outer}}_{\gamma}$) in the
charged curved spacetime. In particular, as suggested in
\cite{Keir}, the presence of this inner {\it stable} null circular
geodesic in the spherically symmetric curved spacetime (\ref{Eq2})
may {\it indicate} that the corresponding horizonless compact object
is non-linearly unstable to massless perturbation fields
\cite{Notekk,Notenw}. One therefore concludes that spatially regular
horizonless spacetimes describing physically realistic ({\it
stable}) compact objects must not possess light rings. This physical
fact yields the lower bound [see Eq. (\ref{Eq20})]
\begin{equation}\label{Eq21}
R>{1\over2}\big(3M+\sqrt{9M^2-8Q^2}\big)\
\end{equation}
on the radii of stable horizonless charged compact objects.

\section{Summary and discussion}

In a physically important paper, Andr\'easson \cite{Hak1} has
recently derived the upper bound (\ref{Eq1}) on the gravitational
masses of spatially regular horizonless charged compact objects. In
the present paper we have raised the physically interesting related
question: Can one derive a stronger upper bound on the gravitational
masses of {\it stable} charged compact systems?

This physically intriguing question is motivated by the recent
theorem presented in \cite{CBH} which, when combined with the
results presented in \cite{Keir}, asserts that horizonless compact
objects whose curved spacetimes possess null circular geodesics
(light rings) are non-linearly unstable to massless perturbation
fields.

Using analytical techniques, we have proved that the answer to the
above stated question is `Yes!'. In particular, it has been
explicitly proved that the masses of physically realistic ({\it
stable}) self-gravitating horizonless compact objects are bounded
from above by the compact functional relation [see Eq. (\ref{Eq21})]
\begin{equation}\label{Eq22}
M<{{R}\over{3}}+{{2Q^2}\over{3R}}\ \ \ \ \ \text{for}\ \ \ \ \ {{Q}\over{M}}\leq \sqrt{{9\over8}}\  .
\end{equation}
Taking cognizance of (\ref{Eq1}) and (\ref{Eq22}) one finds that, in
the dimensionless regime $Q/M\leq \sqrt{9/8}$, the analytically
derived upper bound (\ref{Eq22}) on the gravitational masses of
stable spatially regular charged compact objects is {\it stronger}
than the physically important bound (\ref{Eq1}).
%\cite{Notestro}.
In particular, one finds that the newly derived upper bound
(\ref{Eq22}) is stronger than (\ref{Eq1}) in the $R\geq Q$ regime.
In addition, we recall that in the present paper we consider compact
objects which are characterized by the dimensionless inequality
$Q/M\leq \sqrt{9/8}$ [see (\ref{Eq11})] which, taking cognizance of
Eq. (\ref{Eq21}), corresponds to $R>\sqrt{2}Q$. One therefore
concludes that, in the $Q/M\leq \sqrt{9/8}$ regime, the bound
(\ref{Eq22}) for {\it stable} charged compact systems is stronger
than the bound (\ref{Eq1}).

Finally, it is worth mentioning that a universal upper bound on the
entropies of charged compact systems has been presented in
\cite{Noteunits,BekMay2,Hodebc}:
\begin{equation}\label{Eq23}
S\leq \pi(2MR-Q^2)\  .
\end{equation}
Interestingly, substituting the newly derived upper bound
(\ref{Eq22}) on the masses of physically realistic charged compact
objects into (\ref{Eq23}), one can express the entropy upper bound
in terms of the surface area $A=4\pi R^2$ of the corresponding
charged stable physical system \cite{Noteen}:
\begin{equation}\label{Eq24}
S\leq {{A}\over{6}}-{{\pi Q^2}\over{3}}\  .
\end{equation}

\bigskip
\noindent {\bf ACKNOWLEDGMENTS}
%\bigskip

This research is supported by the Carmel Science Foundation. I would
like to thank Yael Oren, Arbel M. Ongo, Ayelet B. Lata, and Alona B.
Tea for stimulating discussions.


\begin{thebibliography}{99}
\bibitem{Noteunits} We use natural units in which $G=c=\hbar=1$.
\bibitem{Notesth} A null circular geodesic is considered to be
stable (attractive) if massless perturbation fields tend to pile up
on it.
\bibitem{Notekk} This physically intriguing assertion is based on the fact
that massless fields that propagate in the vicinity of unstable
light rings tend to decay over time, whereas they tend to pile up
and grow non-linearly around stable light rings \cite{Keir}.
\bibitem{Notesc} Here $(t,r,\theta,\phi)$ are the Schwarzschild spacetime coordinates.
\bibitem{Noteprm} We use the prime symbol $'$ to denote a
spatial derivative with respect to the radial coordinate $r$.
\bibitem{Noteqq} We shall assume without loss of generality that $Q\geq0$.
\bibitem{Notets} As will be evident below, the external light ring (with $r_{\gamma}>R$) of
the charged compact object exists in the dimensionless bounded
regime (\ref{Eq11}) [see Eq. (\ref{Eq20}) below].
\bibitem{Notethr} Note that the conditions (\ref{Eq12}), which characterize the
null circular geodesics of the spherically symmetric spacetime
(\ref{Eq2}), correspond to $\dot r^2=(\dot r^2)'=0$
\cite{Chan,CarC}.
\bibitem{Notedot} We use the dot symbol $^{.}$ to denote a derivative with respect to an affine parameter.
\bibitem{NoteHoddd} See \cite{Hoddd} for the special case of
horizonless compact objects which are characterized by degenerate
light rings with ${\cal R}={\cal R}'=0$. As discussed in
\cite{Hoddd}, these special matter configurations may possess an odd
number of null circular geodesics.
\bibitem{Notepr} That is, the presence of a null circular geodesic
in the exterior spacetime region with $\rho=p=0$ [see Eq. (\ref{Eq18})].
\bibitem{Notenw} It is important to emphasize that it is a non-trivial
task to prove this instability claim rigorously. Moreover, as
emphasized in \cite{Keir}, even if a spacetime solution of the
Einstein field equations is non-linearly unstable it is difficult to
obtain estimates of the timescale which characterizes this
instability. This implies that an unstable object could still be
observed on finite timescales.
\bibitem{Noteen} It is worth emphasizing again that in the present analysis
we consider horizonless charged objects. On the other hand, black
holes can violate the analytically derived upper bound (\ref{Eq22}),
and as a consequence they can also violate the bound (\ref{Eq24}).
\end{thebibliography}
\end{document}